



\font\twelverm=cmr10 scaled 1200    \font\twelvei=cmmi10 scaled 1200
\font\twelvesy=cmsy10 scaled 1200   \font\twelveex=cmex10 scaled 1200
\font\twelvebf=cmbx10 scaled 1200   \font\twelvesl=cmsl10 scaled 1200
\font\twelvett=cmtt10 scaled 1200   \font\twelveit=cmti10 scaled 1200
\font\twelvesc=cmcsc10 scaled 1200

\skewchar\twelvei='177   \skewchar\twelvesy='60


\def\twelvepoint{\normalbaselineskip=12.4pt
  \abovedisplayskip 12.4pt plus 3pt minus 6pt
  \belowdisplayskip 12.4pt plus 3pt minus 6pt
  \abovedisplayshortskip 0pt plus 3pt
  \belowdisplayshortskip 7.2pt plus 3pt minus 4pt
  \smallskipamount=3.6pt plus1.2pt minus1.2pt
  \medskipamount=7.2pt plus2.4pt minus2.4pt
  \bigskipamount=14.4pt plus4.8pt minus4.8pt
  \def\rm{\fam0\twelverm}          \def\it{\fam\itfam\twelveit}%
  \def\sl{\fam\slfam\twelvesl}     \def\bf{\fam\bffam\twelvebf}%
  \def\mit{\fam 1}                 \def\cal{\fam 2}%
  \def\tt{\twelvett}
  \def\sc{\twelvesc}
  \def\nullspace{\nulldelimiterspace=0pt \mathsurround=0pt }
  \def\big##1{{\hbox{$\left##1\vbox to 10.2pt{}\right.\nullspace$}}}
  \def\Big##1{{\hbox{$\left##1\vbox to 13.8pt{}\right.\nullspace$}}}
  \def\bigg##1{{\hbox{$\left##1\vbox to 17.4pt{}\right.\nullspace$}}}
  \def\Bigg##1{{\hbox{$\left##1\vbox to 21.0pt{}\right.\nullspace$}}}
  \textfont0=\twelverm   \scriptfont0=\tenrm   \scriptscriptfont0=\sevenrm
  \textfont1=\twelvei    \scriptfont1=\teni    \scriptscriptfont1=\seveni
  \textfont2=\twelvesy   \scriptfont2=\tensy   \scriptscriptfont2=\sevensy
  \textfont3=\twelveex   \scriptfont3=\twelveex  \scriptscriptfont3=\twelveex
  \textfont\itfam=\twelveit
  \textfont\slfam=\twelvesl
  \textfont\bffam=\twelvebf \scriptfont\bffam=\tenbf
  \scriptscriptfont\bffam=\sevenbf
  \normalbaselines\rm}



\def\beginlinemode{\endmode
  \begingroup\parskip=0pt \obeylines\def\\{\par}\def\endmode{\par\endgroup}}
\def\beginparmode{\endmode
  \begingroup \def\endmode{\par\endgroup}}
\let\endmode=\par
{\obeylines\gdef\
{}}
\def\singlespace{\baselineskip=\normalbaselineskip}

\def\oneandahalfspace{\baselineskip=\normalbaselineskip
  \multiply\baselineskip by 3 \divide\baselineskip by 2}
\def\doublespace{\baselineskip=\normalbaselineskip \multiply\baselineskip by 2}

\newcount\firstpageno
\firstpageno=2
\footline={\ifnum\pageno<\firstpageno{\hfil}\else{\hfil\twelverm\folio\hfil}\fi}
\let\rawfootnote=\footnote		
\def\footnote#1#2{{\rm\singlespace\hang
  \rawfootnote{#1}{#2\hfill\vrule height 0pt depth 6pt width 0pt}}}

\def\raggedcenter{\leftskip=4em plus 12em \rightskip=\leftskip
  \parindent=0pt \parfillskip=0pt \spaceskip=.3333em \xspaceskip=.5em
  \pretolerance=9999 \tolerance=9999
  \hyphenpenalty=9999 \exhyphenpenalty=9999 }
\def\dateline{\rightline{\ifcase\month\or
  January\or February\or March\or April\or May\or June\or
  July\or August\or September\or October\or November\or December\fi
  \space\number\year}}
\def\received{\vskip 3pt plus 0.2fill
 \centerline{\sl (Received\space\ifcase\month\or
  January\or February\or March\or April\or May\or June\or
  July\or August\or September\or October\or November\or December\fi
  \qquad, \number\year)}}


\hsize=6.5truein
\vsize=8.9truein
\parskip=\medskipamount
\twelvepoint		
\doublespace		
\overfullrule=0pt	


\def\preprintno#1{
 \rightline{\rm #1}}	

\def\title			
  {\null\vskip 3pt plus 0.3fill
   \beginlinemode \doublespace \raggedcenter \bf}

\def\author			
  {\vskip 3pt plus 0.3fill \beginlinemode
   \oneandahalfspace \raggedcenter\sc}

\def\and			
  {\vskip 3pt plus 0.3fill \beginlinemode
   \oneandahalfspace \raggedcenter \rm and}

\def\affil			
  {\vskip 3pt plus 0.1fill \beginlinemode
   \oneandahalfspace \raggedcenter \sl}

\def\abstract			
  {\vskip 3pt plus 0.3fill \beginparmode
   \oneandahalfspace \narrower ABSTRACT:~~}

\def\endtitlepage		
  {\endpage			
   \body}

\def\body			
  {\beginparmode}		

\def\head#1{			
  \filbreak\vskip 0.5truein	
  {\immediate\write16{#1}
   \raggedcenter \uppercase{#1}\par}
   \nobreak\vskip 0.25truein\nobreak}

\def\subhead#1{			
  \vskip 0.25truein		
  {\raggedcenter #1 \par}
   \nobreak\vskip 0.25truein\nobreak}



\def\references
  {\subhead{REFERENCES}
   \frenchspacing \parindent=0pt \leftskip=0.8truecm \rightskip=0truecm
   \parskip=4pt plus 2pt \everypar{\hangindent=\parindent}}

\def\refstylenp{		
  \gdef\refto##1{~[##1]}				
  \gdef\r##1{~[##1]}	         			
  \gdef\refis##1{\indent\hbox to 0pt{\hss[##1]~}}     	
  \gdef\citerange##1##2##3{~[\cite{##1}--\setbox0=\hbox{\cite{##2}}\cite{##3}]}
  \gdef\journal##1, ##2, ##3,                           
    ##4{{##1} {##2} (##3) ##4}}

\def\refstylepr{		
  \gdef\refto##1{$^{##1}$}		
  \gdef\r##1{$^{##1}$}		        
  \gdef\refis##1{\indent\hbox to 0pt{\hss##1.~}}	
  \gdef\citerange##1##2##3{$^\cite{##1}-\setbox0=\hbox{\cite{##2}}\cite{##3}$}
  \gdef\journal##1, ##2, ##3,           		
    ##4.{{##1} {##2}, ##4 (##3).}}

\def\prd{\journal Phys. Rev. D}

\def\prl{\journal Phys. Rev. Lett.}

\def\np{\journal Nucl. Phys.}

\def\pl{\journal Phys. Lett.}

\def\prep{\journal Phys. Rep.}

\def\apj{\journal Astrophys. J.}

\def\endreferences{\body}

\def\figurecaptions		
  {\endpage
   \beginparmode
   \head{Figure Captions}
}

\def\endpage			
  {\vfill\eject}

\def\endpaper			
  {\endmode\vfill\supereject}


\def\ref#1{Ref. #1}			
\def\Ref#1{Ref. #1}			

\def\frac#1#2{{\textstyle{#1 \over #2}}}

\def\sla{\raise.15ex\hbox{$/$}\kern-.57em}
\def\leaderfill{\leaders\hbox to 1em{\hss.\hss}\hfill}
\def\twiddle{\lower.9ex\rlap{$\kern-.1em\scriptstyle\sim$}}
\def\bigtwiddle{\lower1.ex\rlap{$\sim$}}
\def\gtwid{\mathrel{\raise.3ex\hbox{$>$\kern-.75em\lower1ex\hbox{$\sim$}}}}
\def\ltwid{\mathrel{\raise.3ex\hbox{$<$\kern-.75em\lower1ex\hbox{$\sim$}}}}
\def\square{\kern1pt\vbox{\hrule height 1.2pt\hbox{\vrule width 1.2pt\hskip 3pt
   \vbox{\vskip 6pt}\hskip 3pt\vrule width 0.6pt}\hrule height 0.6pt}\kern1pt}
\def\ucsb{Department of Physics\\University of California\\
Santa Barbara, CA 93106}

\def\tev{{\rm \,Te\kern-0.125em V}}
\def\gev{{\rm \,Ge\kern-0.125em V}}
\def\mev{{\rm \,Me\kern-0.125em V}}
\def\kev{{\rm \,ke\kern-0.125em V}}
\def\ev{{\rm \,e\kern-0.125em V}}

\refstylenp


\catcode`@=11
\newcount\r@fcount \r@fcount=0
\newcount\r@fcurr
\immediate\newwrite\reffile
\newif\ifr@ffile\r@ffilefalse
\def\w@rnwrite#1{\ifr@ffile\immediate\write\reffile{#1}\fi\message{#1}}

\def\writer@f#1>>{}
\def\referencefile{
  \r@ffiletrue\immediate\openout\reffile=\jobname.ref%
  \def\writer@f##1>>{\ifr@ffile\immediate\write\reffile%
    {\noexpand\refis{##1} = \csname r@fnum##1\endcsname = %
     \expandafter\expandafter\expandafter\strip@t\expandafter%
     \meaning\csname r@ftext\csname r@fnum##1\endcsname\endcsname}\fi}%
  \def\strip@t##1>>{}}

\def\citeall#1{\xdef#1##1{#1{\noexpand\cite{##1}}}}
\def\cite#1{\each@rg\citer@nge{#1}}	

\def\each@rg#1#2{{\let\thecsname=#1\expandafter\first@rg#2,\end,}}
\def\first@rg#1,{\thecsname{#1}\apply@rg}	
\def\apply@rg#1,{\ifx\end#1\let\next=\relax
\else,\thecsname{#1}\let\next=\apply@rg\fi\next}

\def\citer@nge#1{\citedor@nge#1-\end-}	
\def\citer@ngeat#1\end-{#1}
\def\citedor@nge#1-#2-{\ifx\end#2\r@featspace#1 
  \else\citel@@p{#1}{#2}\citer@ngeat\fi}	
\def\citel@@p#1#2{\ifnum#1>#2{\errmessage{Reference range #1-#2\space is bad.}%
    \errhelp{If you cite a series of references by the notation M-N, then M and
    N must be integers, and N must be greater than or equal to M.}}\else%
 {\count0=#1\count1=#2\advance\count1
by1\relax\expandafter\r@fcite\the\count0,%
  \loop\advance\count0 by1\relax
    \ifnum\count0<\count1,\expandafter\r@fcite\the\count0,%
  \repeat}\fi}

\def\r@featspace#1#2 {\r@fcite#1#2,}	
\def\r@fcite#1,{\ifuncit@d{#1}
    \newr@f{#1}%
    \expandafter\gdef\csname r@ftext\number\r@fcount\endcsname%
                     {\message{Reference #1 to be supplied.}%
                      \writer@f#1>>#1 to be supplied.\par}%
 \fi%
 \csname r@fnum#1\endcsname}
\def\ifuncit@d#1{\expandafter\ifx\csname r@fnum#1\endcsname\relax}%
\def\newr@f#1{\global\advance\r@fcount by1%
    \expandafter\xdef\csname r@fnum#1\endcsname{\number\r@fcount}}

\let\r@fis=\refis			
\def\refis#1#2#3\par{\ifuncit@d{#1}
   \newr@f{#1}%
   \w@rnwrite{Reference #1=\number\r@fcount\space is not cited up to now.}\fi%
  \expandafter\gdef\csname r@ftext\csname r@fnum#1\endcsname\endcsname%
  {\writer@f#1>>#2#3\par}}

\def\ignoreuncited{
   \def\refis##1##2##3\par{\ifuncit@d{##1}%
     \else\expandafter\gdef\csname r@ftext\csname
r@fnum##1\endcsname\endcsname%
     {\writer@f##1>>##2##3\par}\fi}}

\def\r@ferr{\endreferences\errmessage{I was expecting to see
\noexpand\endreferences before now;  I have inserted it here.}}
\let\r@ferences=\references
\def\references{\r@ferences\def\endmode{\r@ferr\par\endgroup}}

\let\endr@ferences=\endreferences
\def\endreferences{\r@fcurr=0
  {\loop\ifnum\r@fcurr<\r@fcount
    \advance\r@fcurr by 1\relax\expandafter\r@fis\expandafter{\number\r@fcurr}%
    \csname r@ftext\number\r@fcurr\endcsname%
  \repeat}\gdef\r@ferr{}\endr@ferences}


\let\r@fend=\endpaper\gdef\endpaper{\ifr@ffile
\immediate\write16{Cross References written on []\jobname.REF.}\fi\r@fend}

\catcode`@=12

\citeall\refto		
\citeall\ref		%
\citeall\Ref		%

\ignoreuncited

\def\e{\eta}
\def\L{\Lambda}
\def\Teq{T_{\rm eq}}
\def\Tosc{T_{\rm osc}}
\def\r{\refto}

\def\gev{{\rm \,Ge\kern-0.125em V}}
\def\mev{{\rm \,Me\kern-0.125em V}}
\def\prep{\journal Phys. Rep.}
\def\apj{\journal Astrophys. J.}

\singlespace

\preprintno{UCSBTH--94--21}
\preprintno{hep-ph/9407323}

\doublespace

\title
CONSTRAINTS ON BARYOGENESIS FROM THE DECAY OF SUPERSTRING AXIONS

\author Raghavan Rangarajan$^{*}$
\affil\ucsb

\abstract
We calculate the dilution of the baryon-to-photon ratio by the decay
of superstring axions.  We find that the dilution is of the order of $10^7$.
We review several models of baryogenesis and show that most of them can not
tolerate such a large dilution.
In particular, only one current model of electroweak  baryogenesis possibly
survives.
The Affleck-Dine mechanism in SUSY GUTs is very robust and the dilution
by axions could contribute to the dilution required in these models.
Baryogenesis scenarios involving topological defects and black hole
evaporation are also capable of
producing a sufficiently large baryon asymmetry.

(*: raghu@tpau.physics.ucsb.edu)
\vskip 1in
{\centerline {\it Nuclear Physics B 454 (1995) 369}}

\endtitlepage

\body
\baselineskip=17.5pt

In a previous paper, hereafter referred to as Paper I\r{r6}, we
used cosmological constraints on the decay of axions that arise in
$E_8^\prime \times E_8$
superstring theories to place a lower limit on the scale of gaugino
condensation in the hidden sector.
In this paper we
calculate
the dilution of the baryon-to-photon ratio
$\eta$ by the decay of superstring axions and use this to constrain
models of baryogenesis.

In $E_8 \times E^\prime_8$ models compactified on a
Calabi-Yau manifold
one typically obtains a model independent axion and several model
dependent axions with decay constants related to the compactification
scale.
If $ E^\prime_8$ breaks down to a non-abelian group there are two
non-abelian groups today: $SU(3)_C$ and the non-abelian subgroup of
$ E^\prime_8$.  The model independent and the model dependent axion degrees
of freedom rearrange themselves to give two physical axions-- the QCD axion
$a$ and the $ E^\prime_8$ axion $a^\prime$\r{choikim85b}.

The mass,
lifetime and energy density of the axion
$a^\prime$
are determined by two energy
scales-- $\Lambda$ and the decay constant $F_a^\prime$.  $F_a^\prime$ is
related to the compactification scale and is about $10^{15} \gev$\r{kim87}.
$\L$ is the scale of gaugino condensation in the hidden sector
which
breaks supersymmetry.
In Paper I, we used limits
from nucleosynthesis, limits on the distortion of the cosmic microwave
and gamma ray backgrounds and closure arguments to show that the $
E^\prime_8$ axion has to decay before 1s, which implies that $\Lambda$
has to be greater than $1.2\times10^{13}\gev$.  $\Lambda$ also
determines the masses of the supersymmetric scalars in the observed
sector $(\tilde m \sim 10^{-1}\L^3/M^2_{Pl})$.  If one assumes that the
supersymmetric scalars have masses of the order of $m_W$ then
$\L$ is approximately $5\times 10^{13}\gev$\r{deribnil}.
It is interesting that our earlier result obtained from astrophysical
and cosmological constraints agrees with the independent
requirement from particle
physics.

In this paper, we let $\L$ equal $5\times 10^{13}\gev$
and study the effects of the decay of the hidden sector axion
$a^\prime$ on the dilution of the baryon-to-photon ratio.
For this value of $\L$, the hidden
sector axion has a mass of $8.7\times 10^5\gev$ and a
lifetime of $2.5\times 10^{-6}$ s.  The axion field starts oscillating
at a temperature of $4.5\times10^{11}\gev$ and dominates the universe at
a temperature of $1.2\times 10^5\gev$ ($1.7 \times 10^{-17}$ s).
Its decay leads to
a dilution of $\e$ by a factor of $9\times 10^6$.  The best lower bound on
$\e$ today,
obtained from upper bounds on the ${\rm D +^3He}$ abundance, gives
$\e_f \ge 3\times 10^{-10}$.  Therefore a successful baryogenesis scenario must
produce a baryon-to-photon ratio of $3\times10^{-3}$ or higher.  Many current
models of baryogenesis are incapable of producing such a large ratio.
In the last section we discuss various models and the baryon asymmetry
that they can attain.

{\centerline{\bf II}}

At high temperatures
the $E_8^\prime$ axion $a^\prime$ is massless.  But as the universe cools it
obtains a potential and a mass $m$.  The axion mass increases till $T \sim \L$
when it attains its final
low-temperature mass $m_0$.  For temperatures below $\L$ the $a^\prime$
mass is given by\refto{choikim85b,dineroseiwit}
$$m_0^2={\pi\over 450}{\Lambda^6\over M^2_{Pl}F_a^{\prime 2}}.
\eqno(1)$$
For $\L$ of $5\times 10^{13}\gev$ and $F_a^\prime$ of $10^{15} \gev$
the low temperature axion mass is $8.7\times10^5\gev$.
In Appendix A
of Paper I
we show that
the universe is radiation dominated when the axion attains its final
low temperature mass at $T\sim \L$.

When the axion potential appears, the axion field need not be at the
minimum of its potential.  At a temperature $T_{\rm osc}$ the field
starts to oscillate about the minimum of the potential with a period
$m^{-1}$.
As we show
in Appendix B
of Paper I,
the axion field starts to oscillate
only after it
attains its low temperature mass, i.e., $T_{\rm osc} < \L$.
The universe is still radiation dominated when the axion field starts
oscillating (Appendix C of Paper I).
Eqn. (8) of Paper I gives
$$T_{\rm osc}=\Biggl[ { m_0 M_{Pl} \over 5 g_{*_{\rm osc}}^{1/2} }\Biggr]^{1/2}
\eqno(2)$$
$g_{*_{\rm osc}}=106.75$
is the effective number of relativistic degrees of freedom
used to calculate energy density.  Therefore $T_{\rm osc}$ is $4.5\times
10^{11} \gev$.
As the universe cools further, it
becomes axion dominated at
a temperature $T_{\rm eq}$.
$T_{\rm eq}$ is $1.2\times 10^5 \gev$
(Appendix A).
Finally, at a temperature $T_{\rm bd}$ the axion decays.
$a^\prime$
decays primarily to two gluons
through the coupling $a^\prime F\tilde {F}$.  The gluons
create jets of mesons
which transfer their energy to the radiation
through scattering and annihilations.

The lifetime of the axion is estimated to be
$$\tau={8200 \pi^5}
{F_a^{\prime 2}\over m_0^3}
\eqno(3)$$
Thus, the lifetime of the axion is $2.5\times 10^{-6}$ s.
In
Appendix A
we show that $t_{\rm eq}$ is $1.7\times 10^{-17}$ s,
i.e., the axion decays only after its energy density
dominates the universe.

We study the zero momentum mode of the $E^\prime_8$
axion field which can be treated
as a condensate
of zero momentum particles.
The axions behave as non-relativistic matter and their energy density
falls as $g_s T^3$.  $g_s$ is the effective number of relativistic
degrees of freedom used to calculate the entropy.
Rewriting the energy density in terms of parameters
at $T_{\rm osc}$ and assuming $A_{\rm osc}\sim F_a^\prime$ (see
Ref.\r{lindeaxion}
for
a
criticism of this assumption) we obtain
$$\rho_{a}=
{5^{3/2}\over2}\Bigl( { g_{s}\over g_{s_{\rm osc}}^{1/4}} \Bigr)
m_0^{1/2} F_a^{\prime 2}
{T^3\over M_{Pl}^{3/2}}.
\eqno(4)$$
$g_{s_{\rm osc}}=g_{*_{ \rm osc}}$.
Further details of the derivation of the energy density and of the
cosmology of superstring axions are given in
Paper I.

{\centerline{\bf III}}

Now we calculate the dilution of the baryon asymmetry of
the universe due to the decay of the
axions.  We use the approximation that all decays occur at $t=\tau$ and
that the resulting radiation heats up the universe at a temperature
$T_{\rm bd}$ to a temperature
$T_{\rm ad}$.  Scherrer and Turner\r{schturn85}
have shown that if one takes into
consideration the exponential decay of massive particles, the resulting
radiation does not heat up the universe; it causes the universe to cool more
slowly to the temperature  $T_{\rm ad}$.
However, as they point out,
their calculation of the change in the entropy is not very different
from ours using the naive approximation of instantaneous decay.

The baryon-to-photon ratio $\e$ is defined as $n_B/n_\gamma$, where
$n_B$ is the baryon number density.  We shall assume that
the baryon number does not change after baryogenesis.  (Any baryon number
violating interactions after baryogenesis
will only decrease $n_B$ and strengthen our result.)
Therefore,
$$n_{B_i}/n_{B_f}=R_f^3/R_i^3,
\eqno(5)$$
where $i$ and $f$ refer to values at the time of baryogenesis and today
respectively and $R$ is the scale factor.
$$n_{\gamma_f}/n_{\gamma_i}={T_f^3\over T_i^3}={S_f\over g_{s_f} R_f^3}
{ g_{s_i} R_i^3\over S_i}
\eqno(6)$$
where $S$ is the entropy.  $g_{s_i}$ and $g_{s_f}$ are 106.75 and 3.9
respectively.
The entropy of the universe does not change when a species disappears
through annihilations in equilibrium (such as for $e^\pm$ annihilations).
Therefore, $S_i=S_{\rm bd}$ and $S_f=S_{\rm ad}$, where `bd'
and `ad' refer to
`before decay' and `after decay' respectively.
$${S_{\rm ad}\over S_{\rm bd}}={g_{s_{\rm ad}} T_{\rm ad}^3 R_{\rm ad}^3 \over
g_{s_{\rm bd}} T_{\rm bd}^3 R_{\rm bd}^3 }
\eqno(7)$$
In the instantaneous decay approximation $R_{\rm bd}=R_{\rm ad}$ and since the
axion is non-relativistic when it decays $g_{s_{\rm bd}}=g_{s_{\rm ad}}=70$.
Thus,
$${\e_i\over \e_f}= {g_{s_i} T_{\rm ad}^3\over g_{s_f} T_{\rm bd}^3}
\eqno(8)$$

{}From conservation of energy we get
$$T_{\rm ad}^3=\Biggl [ T_{\rm bd}^4 +
\rho_{a_{\rm bd}} {30\over \pi^2 g_{*_{\rm ad}}}
\Biggr ]^{3/4}
\eqno(9)$$
To obtain $T_{\rm bd}$ we first derive the time-temperature relation
for $t_{\rm eq}<t<\tau$.  For an $\Omega=1$ universe,
$$H=\Biggl ({8 \pi\over 3 M_{Pl}^2} \rho \Biggr )^{1/2}
\eqno(10)$$
When the universe is axion dominated we use the axion energy density
$\rho_a$ in
(4).
Also, for $t\gg t_{\rm eq}, H=(2/3t)$.
Setting $t=\tau$ we get
$$T_{\rm bd}=0.21\Biggl[{g_{s_{\rm osc}}^{1/4} \over g_{s_{\rm bd}}}
 {M_{Pl}^{7/2} \over \tau^2 m_0^{1/2} F_a^{\prime 2}}
\Biggr ]^{1/3}
\eqno(11)$$
We also get $\rho_{a_{\rm bd}}$ from
(10).
$$\rho_{a_{\rm bd}}={M_{Pl}^2 \over 6\pi \tau^2}
\eqno(12)$$
Thus $T_{\rm bd}$ is $5.8\mev$ and $T_{\rm ad}$ is $0.39\gev$.

Substituting (9), (11) and (12) in (8) we get
$${\e_i\over \e_f}={g_{s_i}\over g_{s_f}}
\Biggl[ 1+ \Bigl( {M_{Pl}^2\over 6 \pi\tau^2} \Bigr)
\Bigl( {30\over \pi^2 g_{*_{\rm ad}}}\Bigr)
\Biggl( 110
{ g_{s_{\rm bd}} \over g_{s_{\rm osc}}^{1/4}}
{\tau^2 m_0^{1/2}F_a^{\prime 2}\over M_{Pl}^{7/2}}\Biggr)^{(4/3)}\Biggr]^{3/4}
\eqno(13)$$
Since the second term in the bracket is much larger than 1, or equivalently,
$\rho_{a_{\rm bd}} \gg \rho_{\rm rad}$ at the time of the decay
$${\e_i\over \e_f}=28{g_{s_i}\over g_{s_f}}
{ g_{s_{\rm bd}} \over g_{s_{\rm osc}}^{1/4} g_{*_{\rm ad}}^{3/4}}
{\tau^{1/2} m_0^{1/2} F_a^{\prime 2}\over M_{Pl}^{2}}
\eqno(14)$$
Substituting for $\tau$ and $m_0$ in terms of $\L$ we get
$${\e_i\over \e_f}=5.3\times 10^5 {g_{s_i}\over g_{s_f}}
{ g_{s_{\rm bd}} \over g_{s_{\rm osc}}^{1/4} g_{*_{\rm ad}}^{3/4}}
{F_a^{\prime 4}\over \L^3 M_{Pl}}
\eqno(15)$$
For $\L$ of $5\times 10^{13}\gev$ and $F_a^{\prime}$ of $10^{15}\gev$,
the dilution is
$${\e_i\over \e_f}=9\times 10^6
\eqno(16)$$
For a conservative lower bound on $\e_f$ of $3\times 10^{-10}$, we find
that $\e_i$ must be greater than $3\times 10^{-3}$
or $n_B/s$, the baryon-to-entropy ratio, must be greater than $10^{-5}$.
$n_B/s$ today is $4\times10^{-11}$.

\vfill\eject

{\centerline{\bf IV}}

{\bf Discussion:}

The simplest mechanism for producing a baryon asymmetry involves
out-of-equilibrium decays of massive Higgs or gauge bosons in GUTs
or supersymmetric GUTs.  This mechanism can, in principle, produce a
baryon-to-entropy ratio of $10^{-4}$\r{kolbturner}.  However,
the maximum asymmetry obtained in most specific GUT scenarios is the
observed asymmetry today or less\r{dolgov}.
In general, one obtains an even smaller asymmetry in supersymmetric
GUTs.
In the context of an inflationary universe, out-of-equilibrium decays in
GUTs fare worse\r{cdo93b}.
The COBE results constrain the inflationary potential
and prescribes a low inflaton mass$\sim 10^{11}\gev$.  The out-of-equilibrium
decays scenarios then require the massive boson to be lighter than the inflaton
which makes it difficult to obtain even the present asymmetry.  Again, the
situation is much worse for SUSY GUTs.

However, the Affleck-Dine mechanism involving the decay of sfermion condensates
in SUSY GUTs in an inflationary universe
can produce a baryon-to-entropy ratio as large as
$10^{-2}$\r{affledinelinde}.
Furthermore, Davidson et al.\r{davimurayoliv} have shown that
the presence of Bose-Einstein condensates can suppress the
destruction of the baryon
asymmetry by electroweak sphaleron processes.
To preserve a Bose-Einstein condensate till the electroweak phase transition
in an inflationary universe,
below which sphaleron processes are naturally supressed,
requires $n_B/s\ge0.01$.  It will subsequently need to be
diluted to about
$4\times 10^{-11}$ before
nucleosynthesis.  Superstring axion decays can certainly
provide part of the required
dilution.
(For other mechanisms to dilute the entropy see
ref.\r{yametal}.)

Fukugita and Yanagida\r{fy} (also see ref.\r{luty}), Lazarides and
Shafi\r{lazshaf91} and
Campbell et al.\r{cdo93a,cdo93b}
produce a baryon asymmetry by first creating a lepton asymmetry and then
convert this into a baryon asymmetry by sphaleron proccesses.  In the
Fukugita-Yanagida and Lazarides-Shafi
mechanism, a lepton asymmerty is created by the
out-of-equilibrium
decay of heavy right-handed Majorana neutrinos in a see-saw
mechanism.  In the model of Lazarides and Shafi the heavy neutrinos are
obtained by the decay of the inflaton field.  Campbell et al.\r{cdo93b}
consider a supersymmetric version of the Fukugita-Yanagida model
and also study it within the context of an inflationary
universe.  Though these models can produce the asymmetry observed today
they can not create an asymmetry as high as $10^{-5}$.  Campbell
et al. also create a lepton asymmetry by the effect of lepton number
violating induced operators, arising from see-saw (s)neutrino masses,
which act on scalar condensate oscillations along flat directions of the
supersymmetric standard model.  As in the Affleck-Dine mechanism, a
large asymmetry of $10^{-5}$ can be created.

In recent years, there have been a number of models attempting to create the
baryon asymmetry at the electroweak transition, generally in a first-order
phase transition.  In a model by Nelson et al.\r{nkc92}, a hypercharge
asymmetry is created outside a bubble of true vacuum by a difference in the
reflection rates for left- and right-handed top quarks
bouncing off the  bubble wall in the false vacuum.
The hypercharge asymmetry is converted into a baryon asymmetry.  They can
obtain
a baryon-to-entropy ratio of $10^{-6}$.  However that assumes maximal CP
violation which is not very likely\r{kap}.
In a similar scenario, a lepton asymmetry is created by the reflection of heavy
neutrinos off the bubble wall\r{ckn90cknel91}.
The lepton asymmetry is then converted into a
baryon asymmetry by B+L violating processes.  This scenario can produce
the required $n_B/s$ of $10^{-5}$
if there is a fourth neutrino heavier than $45\gev$
(as required by the data from Z decays) and the
neutrino mixing angles are small.
However this value may be diminished by diffusion of particles from the
true vacuum to the false vacuum\r{maalampi94}.
Furthermore, one has to check that the heavy neutrino spends
enough time in the symmetric phase after reflection for sphaleron
processes to convert the lepton asymmetry into a baryon asymmetry,
before the wall catches up with it\r{kap}.

Turok and Zadrozny\r{turokz9091}
and McLerran et al.\r{mclv91}
produce a baryon asymmetry through the CP asymmetric interaction
of the wall with field fluctuations of non-zero Chern-Simons number
in the the symmetric phase\r{ambj8789}.
Sufficient CP violation is typically provided by extending the Standard Model
to two or more Higgs doublets.  The above scenarios can
produce a baryon asymmetry that is close to the present value only.
(Dine et al.\r{dinehues92} have indicated that the baryon asymmetry
produced will be even lower.)

Another mechanism for producing the baryon asymmetry at the
electroweak phase transition has been proposed by McLerran\r{mclerran}.
In this model CP violation is provided by the existence of a QCD axion.
Interference between electroweak sphaleron-induced baryon number violating
processes and QCD sphaleron-induced CP violating processes (which require
$\theta_{\rm QCD}\ne 0$) produces a baryon asymmetry.  In the standard model
this scenario can not produce the present day asymmetry.  Introducing heavy
scalars in the theory can give a maximum $n_B/s$ of $10^{-8}$.

Theories of spontaneous baryogenesis involve the spontaneous breaking
of a $U(1)_B$ baryon symmetry\r{cohkap8788}.
Regions of baryons and anti-baryons, inflated
to supra-horizon sizes, are produced by the slow rolling of the consequent
pseudo-Goldstone boson towards the minimum of its potential and by its decay
as it finally oscillates about its minimum.  While such scenarios can produce
the observed asymmetry they can not tolerate a dilution of the order of
$10^5$.  This idea of creating a baryon asymmetry by using
the evolution of a scalar to create an effective chemical potential
has been extended to the electroweak phase transition by
Dine et al.\r{dinhuesingsuss,dinehues92} and
by Cohen et al.\r{ckn91} and Dine and Thomas\r{dinethomas94}.
The maximum baryon-to-entropy ratio they
obtain is of the order of $10^{-7}$.

Topological defects have also been used to create a baryon asymmetry.
Nussinov\r{nuss} has considered monopole-anti-monopole annihilations as
a source of heavy GUT Higgs and gauge bosons which decay asymmetrically
into baryons and anti-baryons giving a baryon asymmetry.  This model can
produce a maximum baryon-to-entropy ratio of $10^{-10}$.  Bhattacharjee
et al.\r{bhatt82} can obtain a baryon asymmetry of $10^{-5}$ from
the decay of heavy Higgs and gauge bosons produced by
collapsing cosmic string loops if the GUT scale is about $10^{16}\gev$.
Brandenberger et al.\r{brand91} have also obtained a high baryon
asymmetry similarly from collapsing string loops.  Kawasaki and Maeda\r{kawama}
consider a baryon asymmetry created by the decay of particles emitted
from cusps moving close to the speed of light on cosmic strings, as well
as from kinky string loops, proposed by Garfinkle and
Vachaspati\r{garvach87}.  They can create a large enough asymmetry but the
existence of cusp evaporation is uncertain.  Mohazzab\r{mohazzab}
also obtains a baryon asymmetry as high as $10^{-5}$ from cusp
annihilation on long cosmic strings.
Barrow et al.\r{barrow91a} create a baryon asymmetry from the decay of bosons
produced
in bubble wall collisions in extended inflation models.  They can obtain
a baryon asymmetry of $10^{-5}$ only if the colliding bubbles are not much
larger than the critical radius.

Several models have been proposed in which a baryon asymmetry is created
by the evaporations of black holes.  Zel'dovich\r{zeldovich} and
Dolgov\r{dolgov81} showed that one can obtain a baryon asymmetry from
black holes even if baryonic charge is conserved.  Though initially the
black hole emits particles that decay to
an equal number of baryons and antibaryons, more
antibaryons than baryons are recaptured by the black hole.  Toussaint et
al.\r{tous79}, Turner and Schramm\r{turnerschramm},
Turner\r{turner79}, Barrow\r{barrow80} and Barrow and Ross\r{barrowross}
have suggested that black holes can emit heavy bosons and produce a
baryon asymmetry by their decay.  These models require baryon number
violation and are similar to the GUT baryogenesis scenarios. Barrow et
al.\r{barrow91b} have obtained a baryon asymmetry from the evaporation
of black holes formed during extended inflation.  Black hole evaporation
can give a very large baryon asymmetry.  However it is difficult to
estimate their number density and mass distribution.  Hence there is
large uncertanty in the baryon asymmetry that can actually be created.

{\bf Conclusion:}
Thus we see that the existence and decay of superstring axions is not
compatible with many models of baryogenesis.  Only one current model of
electroweak baryogenesis may be able to
tolerate the dilution of the baryon asymmetry.
The Affleck-Dine mechanism
is certainly very robust. Models involving topological defects
and black hole evaporation are
also capable of creating a large enough baryon asymmetry.

{\bf Acknowledgements:}
I would like to thank Mark Srednicki, Subir Sarkar, Robert Scherrer, David
Kaplan and
Kiwoon Choi for very useful
discussions.  I would also like to thank the referee for pointing out
the possibility of the induced decay of axions due to resonance
effects.
I am also grateful to the Center for Particle Astrophysics
at the University of California, Berkeley, where
part of this work was completed, for their hospitality.

After this work was completed, I discovered that similar work on the
cosmological consequences of scalars in superstring and supergravity
theories has been carried
out by other authors\r{earlier}.

This work was supported by NSF Grant No. PHY91-16964.

\endpage

\centerline{\bf Appendix A}

In this Appendix we show that the axion decays after its energy density
dominates the universe.

At $t=t_{\rm eq}$,
$$(4t_{\rm eq}^2)^{-1}={8\pi\over 3 M_{Pl}^2}
\Biggl ( {1\over 2}m_0^2 A^2(T_{\rm eq})\Biggr)
\eqno(A.1)$$
Also,
$$m_0 A^2(T_{\rm eq})=m_0 A^2(T_{\rm osc}){T_{\rm eq}^3\over T_{\rm osc}^3}
\eqno(A.2)$$
where we have set $g_{s_{\rm osc}}=g_{s_{\rm eq}}=106.75$.  $T_{\rm osc}$ is
given by
(2).
We now calculate $T_{\rm eq}$.

At $T_{\rm eq}$,
$$\eqalignno{ {\pi^2\over 30} g_{*_{\rm eq}}T_{\rm eq}^4&=
\Biggl ( {1\over 2}m_0^2 A^2(T_{\rm eq})\Biggr)\cr &=
\Biggl ({\pi\over 900} {\L^6\over M_{Pl}^2 F_a^{\prime 2}} \Biggr )
A^2(T_{\rm osc}){T_{\rm eq}^3\over T_{\rm osc}^3}
&(A.3)\cr }$$
Therefore,
$$\eqalignno {\Teq&={1.1\times 10^{-2}\over g_{*_{\rm eq}}}
{\L^6\over M_{Pl}^2 \Tosc^3} &(A.4)\cr
&=1.5 {\L^{3/2}F_a^{\prime 3/2}\over M_{Pl}^2} &(A.5)\cr}$$
$\Teq=1.2\times10^5\gev$.

Combining (A.1), (A.2), (A.5) and
(2),
we  get
$$\eqalignno{t_{\rm eq}&=
1.3\times 10^{-2}  {M_{Pl}^5\over \L^3 F^{\prime 3}_a}\cr
&= 1.7\times 10^{-17}{\rm s}&(A.6)\cr}$$
However $\tau=8200\pi^5 (F_a^{\prime 2}/m_0^3)$ gives a lifetime of $2.5
\times 10^{-6}$ s.  Thus the axion decays after it has dominated the universe,
or $\Teq>T_{\rm bd}$.

\references

\hyphenation{Amsterdam}


\refis{ranga94a}R. Rangarajan, \np, B454, 1995, 369.

\refis{r6}R. Rangarajan, \np, B454, 1995, 357.


\refis{pww83}J. Preskill, M. B. Wise, and F. Wilczek,
\pl, B120, 1983, 127.

\refis{as83}L. Abbott and P. Sikivie, \pl, B120, 1983, 133.

\refis{df83}M. Dine and W. Fischler, \pl, B120, 1983, 137.

\refis{lindeaxion}A. D. Linde, \pl, B201, 1988, 437; \pl, B259, 1991, 38.

\refis{srednicki90}Particle Physics and Cosmology: Dark Matter,
ed. M. Srednicki (North--Holland, Amsterdam, 1990).

\refis{turner86}M. S. Turner, \prd, 33, 1986, 889.  

\refis{witten85}E. Witten, \pl, B155, 1985, 151. 

\refis{linde83}A. D. Linde, \pl, 129B, 1983, 177.



\refis{ghmr}D. J. Gross, J. A. Harvey, E. Martinec and R. Rohm,
\prl, 54, 1985, 502.

\refis{greenschwartz}M. B. Green and J. H. Schwarz, \pl, B149,
1984, 117.

\refis{ghmrgreenschwartz}D. J. Gross, J. A. Harvey, E. Martinec and R. Rohm,
{\it Phys. Rev. Lett.} {\bf 54}, (1985), 502;
M. B. Green and J. H. Schwarz,
{\it Phys. Lett.} {\bf B149}, (1984), 117.

\refis{choikim85a}K. Choi and J. E. Kim, \pl, B154, 1985, 393.

\refis{choikim85b}K. Choi and J. E. Kim, \pl, B165, 1985, 71.

\refis{kim87}J. E. Kim, \prep, 150, 1987, 1.

\refis{sred85}M. Srednicki,
\np, B260, 1985, 689.

\refis{kap85}D. Kaplan,
\np, B260, 1985, 215.

\refis{geokapran86}H. Georgi, D. Kaplan and L. Randall,
\pl, B169, 1986, 73.

\refis{deribnil}J. P. Derendinger, L. E. Ibanez and H. P. Nilles,
\pl, B155, 1985, 65.

\refis{dineroseiwit}M.Dine, R. Rohm, N. Seiberg and E. Witten, \pl,
B156, 1985, 55.

\refis{barrchoikim}S. M. Barr, K. Choi, and J. E. Kim, \np, B283, 1987,
591.

\refis{barr85}S. M. Barr, \pl, B158, 1985, 397.

\refis{wenwit}X. G. Wen and E. Witten, \pl, B166, 1986, 397.

\refis{dsww}M. Dine, N. Seiberg, X. G. Wen and E. Witten, \np, B289,
1987, 319.

\refis{wenwitdsww}X. G. Wen and E. Witten,
{\it Phys. Lett.} {\bf B166} (1986) 397;
M. Dine, N. Seiberg, X. G. Wen and E. Witten, {\it Nucl. Phys.}
{\bf B289} (1987) 319.

\refis{witten84}E. Witten, \pl, B149, 1984, 351.

\refis{witten85}E. Witten, \pl, B153, 1985, 243.

\refis{witten85b}E. Witten, \pl, B155, 1985, 151.

\refis{canhorstrowit}P. Candelas, G. T. Horowitz, A. Strominger and
E. Witten, \np, B258, 1985, 46.

\refis{lindebk}A. D. Linde, Particle Physics and Inflationary Cosmology,
(Harwood Academic Publishers, Switzerland, 1990).

\refis{pwwasdf83}J. Preskill, M. B. Wise, and F. Wilczek,
{\it Phys. Lett.}, {\bf B120} (1983) 127;
L. Abbott and P. Sikivie, {\it ibid.}, 134;
M. Dine and W. Fischler, {\it ibid.}, 137.

\refis{schertur88b}R. J. Scherrer and M. S. Turner, \apj, 331, 1988, 33.

\refis{kolbt}E. W. Kolb and M. S. Turner, The Early Universe
(Addison-Wesley Publishing Company, 1990).

\refis{schturn85}R. J. Scherrer and M. S. Turner, \prd, 31, 1985, 681.

\refis{kolbturner}E. W. Kolb and M. S. Turner, \journal
Ann. Rev. Nucl. Part. Sci., 33, 1983, 645.

\refis{dolgov}A. D. Dolgov, \prep, 222, 1992, 309.
This is an excellent review of baryogenesis models.

\refis{davimurayoliv}S. Davidson, H. Murayama and K. A. Olive,
hep-ph 9403259, (1994).

\refis{cdo93b}B. Campbell, S. Davidson and K. A. Olive, \np, B399, 1993,
111.

\refis{affledine}I. Affleck and M. Dine, \np, B249, 1985, 361.

\refis{linde}A. D. Linde, \pl, B160, 1985, 243.

\refis{affledinelinde}I. Affleck and M. Dine, {Nucl. Phys.}
{B249}, (1985)
361; A. D. Linde, {Phys. Lett.} {B160}, (1985) 243.

\refis{dolkiri}A. D. Dolgov and D. P. Kirilova, \journal Sov. J. Nucl. Phys,
50, 1989, 1006.

\refis{yamamoto}K. Yamamoto, \pl, B168, 1986, 341.

\refis{enqnanquiros}K. Enqvist, D. V. Nanapoulos and M. Quiros, \pl,
B169, 1986, 343.

\refis{ellennanoliv}J. Ellis, K. Enqvist, D. V. Nanopoulos and K. A.
Olive, \pl, B188, 1987, 415.

\refis{yametal}K. Yamamoto, {Phys. Lett.} {B168}, (1986) 341;
K. Enqvist, D. V. Nanapoulos and M. Quiros, {Phys. Lett.}
{B169}, (1986) 343; J. Ellis, K. Enqvist, D. V. Nanopoulos and K. A.
Olive; {Phys. Lett.} {B188}, (1987) 415.

\refis{nkc92}A. E. Nelson, D. B. Kaplan and A. G. Cohen, \np, B373,
1992, 453.

\refis{kap}D. B. Kaplan, private communication.

\refis{cknel91}A. G. Cohen, D. B. Kaplan and A. E. Nelson, \np, B349,
1991, 727.

\refis{ckn90}A. G. Cohen, D. B. Kaplan and A. E. Nelson, \pl, B245,
1990, 561.

\refis{ckn90cknel91}A. G. Cohen, D. B. Kaplan and A. E. Nelson, \pl, B245,
1990, 561; \np, B349, 1991, 727.

\refis{maalampi94}J. Maalampi, J. Sirrka and I. Vilja, hep-ph 9405378,
(1994).

\refis{ambj87}J. Ambjorn, M. Laursen and M. E. Shaposhnikov, \pl,
B197, 1987, 49.

\refis{ambj89}J. Ambjorn, M. Laursen and M. E. Shaposhnikov, \np,
B316, 1989, 483.

\refis{ambj8789}J. Ambjorn, M. Laursen and M. E. Shaposhnikov, \pl,
B197, 1987, 49; \np, B316, 1989, 483.

\refis{turokz90}N. Turok and J. Zadrozny, \prl, 65, 1990, 2331.

\refis{turokz91}N. Turok and J. Zadrozny, \np, B358, 1991, 471.

\refis{turokz92}N. Turok and J. Zadrozny, \np, B369, 1992, 729.

\refis{turokz9091}N. Turok and J. Zadrozny, \prl, 65, 1990, 2331;
\np, B358, 1991, 471.

\refis{mclv91}L. McLerran, M. Shaposhnikov, N. Turok and M. Voloshin,
\pl, B256, 1991, 451.

\refis{dinehues92}M. Dine, P. Huet and R. Singleton Jr.,
\np, B375, 1992, 625.

\refis{mclerran}L. McLerran, \prl, 62, 1989, 1075.

\refis{cohkap88}A. G. Cohen and D. B. Kaplan, \np, B308, 1988, 913.

\refis{cohkap87}A. G. Cohen and D. B. Kaplan, \pl, B199, 1987, 251.

\refis{cohkap8788}A. G. Cohen and D. B. Kaplan, \pl, B199, 1987, 251;
\np, B308, 1988, 913.

\refis{dinhuesingsuss}M. Dine, P. Huet, R. Singleton Jr. and L. Susskind,
\pl, B257, 1991, 351.


\refis{ckn91}A. G. Cohen, D. B. Kaplan and A. E. Nelson, \pl,
B263, 1991, 86; hep-ph 9406345, (1994).

\refis{dinethomas94}M. Dine and S. Thomas, hep-ph 9401265, (1994).

\refis{fy}M. Fukugita and T. Yanagida, \pl, B174, 1986, 45;
\prd, 42, 1990, 1285.

\refis{lazshaf91}G. Lazarides and Q. Shafi, \pl, B258, 1991, 305.

\refis{luty}M. Luty, \prd, 45, 1992, 455.

\refis{cdo93a}B. Campbell, S. Davidson and K. A. Olive, \pl, B303, 1993,
63.

\refis{nuss}S. Nussinov, \pl, B110, 1982, 221.

\refis{bhatt82}R. Bhattacharjee, T. W. Kibble and N. Turok, \pl, B119,
1982, 95.

\refis{brand91}R. H. Brandenberger, A. C. Davis and M. Hindmarsh, \pl,
B263, 1991, 239.

\refis{kawama}M. Kawasaki and K. Maeda, \pl, B208, 1988, 84.

\refis{garvach87}D. Garfinkle and T. Vachaspati, \prd, 36, 1987, 2229.

\refis{mohazzab}M. Mohazzab, hep-ph 9409274, (1994).

\refis{holdom83}B. Holdom, \prd, 28, 1983, 1419.

\refis{barrow91a}J. D. Barrow, E. J. Copeland, E. W. Kolb and A. R.
Liddle, \prd, 43, 1991, 977.

\refis{zeldovich}Ya. B. Zel'dovich, \journal JETP Lett., 24, 1976, 24.

\refis{dolgov81}A. D. Dolgov, \prd, 24, 1981, 1042.

\refis{tous79}D. Toussaint, S. B. Treiman, F. Wilczek, A. Zee,
\prd, 19, 1979, 1036.

\refis{turnerschramm}M. S. Turner and D. N. Schramm, \journal Nature,
279, 1979, 303.

\refis{turner79}M. S. Turner, \pl, B89, 1979, 155.

\refis{barrow80}J. D. Barrow,  \journal Mon. Not. R. Astr. Soc., 192,
1980, 427.

\refis{barrowross}J. D. Barrow  and G. G. Ross, \np, B181, 1981, 461.

\refis{barrow91b}J. D. Barrow, E. J. Copeland, E. W. Kolb and
A. R. Liddle, \prd, 43, 1991, 984.

\refis{earlier}G. D. Coughlan, W. Fischler, E. W. Kolb, S. Raby and
G. G. Ross,  {Phys. Lett.} {B131} (1983) 59;
A. S. Goncharov, A. D. Linde and
M. I. Vysotsky, {Phys. Lett.} {B147} (1984) 279;
G. German and G. G. Ross, {Phys. Lett.} {B172} (1986)
305;
J. Ellis, D. Nanopoulos and M. Quiros, {Phys. Lett.} {B174}
(1986) 176;
O. Bertolami, {Phys. Lett.} {B209} (1988) 277.

\endreferences
\end